\documentclass[aps,prl,twocolumn,superscriptaddress,floatfix]{revtex4-2}
\usepackage{multirow}
\usepackage[latin9]{inputenc}
\usepackage{times}
\usepackage{amssymb}
\usepackage{amsmath}
\usepackage{graphicx}
\usepackage{graphics}
\usepackage{color}
\usepackage{bm}

\makeatletter

\newcommand{\mfA}{\mathfrak{A}}

\newcommand{\cA}{{\cal A}}
\newcommand{\cV}{{\cal V}}

\newcommand{\rev}[1]{\textcolor{black}{{#1}}}
\newcommand{\rerev}[1]{\textcolor{black}{{#1}}}

\begin{document}

\title{
Nonlinear Thouless pumping: solitons and transport breakdown}

\author{Qidong Fu}
\affiliation{School of Physics and Astronomy, Shanghai Jiao Tong University, Shanghai 200240, China}

\author{Peng Wang}
\affiliation{School of Physics and Astronomy, Shanghai Jiao Tong University, Shanghai 200240, China}

\author{Yaroslav V. Kartashov}
\affiliation{Institute of Spectroscopy, Russian Academy of Sciences, Troitsk, Moscow Region, 108840, Russia}

\author{Vladimir V. Konotop}
\affiliation{Departamento de F\'{i}sica and Centro de F\'{i}sica Te\'orica e Computacional, Faculdade de Ci\^encias, Universidade de Lisboa, Campo Grande, Ed. C8, Lisboa 1749-016, Portugal}
 
\author{Fangwei Ye}
\email[]{Corresponding author: fangweiye@sjtu.edu.cn}
\affiliation{School of Physics and Astronomy, Shanghai Jiao Tong University, Shanghai 200240, China}

\begin{abstract}
One-dimensional topological pumping of matter waves in two overlaid optical lattices moving with respect to each other is considered in the presence of attractive nonlinearity. It is shown that there exists a threshold nonlinearity level above which the matter transfer is completely arrested. Below this threshold the transfer of both dispersive wavepackets and solitons occurs in accordance with the predictions of the linear theory, i.e. it is quantized and determined by the dynamical Chern numbers of the lowest band. The breakdown of the transport is also explained by nontrivial topology of the bands. In that case, the nonlinearity induces Rabi oscillations of atoms between two (or more) lowest bands. If the sum of the dynamical Chern numbers of the populated bands is zero, the oscillatory dynamics of a matter soliton in space occurs, which corresponds to the transport breakdown. Otherwise the sum of the Chern numbers of the nonlinearity-excited bands determines the direction and magnitude of the average velocity of matter solitons that remains quantized and admits fractional values. \rerev{Thus, even in strongly nonlinear regime the topology of the linear bands is responsible for the evolution of solitons.} The transition between different dynamical regimes is accurately described by the perturbation theory for solitons.

\end{abstract}

\maketitle

Controlled unidirectional transfer of matter, heat, or physical characteristics like charge or spin in a medium is a fundamental problem.
It was discovered by Thouless~\cite{Thouless} that one-dimensional (1D) pumping of electrons by a slowly moving periodic potential over one pumping cycle is quantized with quanta being determined by the Chern numbers considered in the extended coordinate-time space. To date Thouless pumping was experimentally observed in systems of cold bosonic~\cite{LosZilAlBlo2020,LoSHa2018,NaTaKe2021} and fermionic~\cite{NaToTa2016} atoms,  spin systems~\cite{spin},  optics~\cite{KraLaRin2012,ZilHuaJo2018,CeWaShe2020}, acoustics~\cite{CheProPro2020}, and plasmonics~\cite{FedQiLIK2020}.

Although topological pumping was originally introduced for essentially linear systems~\cite{Thouless,MiChaNiu2010}, physical realizations mentioned above provide opportunities for investigation of the phenomenon in nonlinear settings, for example, in the presence of optical Kerr nonlinearity or two-body interactions in Bose-Einstein condensates (BECs). Inter-atomic interactions were intrinsically present in experiments with BECs in bipartite magnetic lattices where (non-quantized) transport was observed~\cite{LuSAG2016} and in the formation of Mott insulator phase assuring tight binding of atoms upon pumping~\cite{LoSHa2018}. Nonlinear effects were explicitly included in the single-band tight-binding approximation for a gas of several spinor states of fermions in optical lattices~\cite{TaCoFa2017} and in model of a non-linear interferometer~\cite{HaDuKAm2019}, where the obtained pump was fractional. Topological transport of interacting photons was described in~\cite{TanDasAn2016}. Qualitatively new effects appear in the presence of sufficiently strong interactions. Using a Rice-Mele model, accounting for spin and Hubbard interaction term, in~\cite{NaYoKa2018} the authors reported the breakdown of the Thouless pumping, explained by closure of the gap computed using amplitude-dependent many-body ground states under twisted boundary conditions. Apparently, the effect may be dependent on the specific model, since in a another realization of a spinful Rice-Mele model subject to both open and twisted boundary conditions, breakdown of pumping was not found~\cite{SteHaHe2019}.

All above results dealt with repulsive inter-atomic interactions, and thus with stable nonlinear states of constant amplitudes. In this Letter we address  nonlinear Thouless pumping of matter waves in a BEC with a negative scattering length, loaded in an optical superlattice created by two lattices, one moving with respect to the other. In this regime a well-localized matter soliton is formed and
  excitations from the upper bands of the linear lattice spectrum acquire special importance~\cite{AKKS,BraKon2004}. Our two main findings are: First, there exists a threshold amplitude below which the pumping closely follows the prediction of the linear theory~\cite{Thouless}, even when solitons are formed. Above the threshold amplitude, the transport either becomes fractional,  or acquires direction opposite to the moving sub-lattice velocity, or is arrested resulting in oscillatory motion of a soliton. Second, oscillations of a 
soliton, relatively well described by the perturbation theory~\cite{perturb}, have topological nature. They occur due to the nonlinearity-induced Rabi oscillations of atoms between bands with different Chern numbers. The sum of the Chern numbers of the excited bands determines the direction and fractional magnitude of the one-cycle-average velocity of the soliton. While dynamics of nonlinear wavepackets in oscillating lattices was studied previously in discrete models~\cite{KCV,BisSal}, in photonics~\cite{optlatt01,optlatt02,optlatt03,zhangoe2015}, and in BECs~\cite{BEC1,BEC2,StaLon}, the \rev{relation of the transport properties of nonlinear systems with the topological properties of linear bands is established here for the first time}.

The meanfield dynamics of 1D BEC \rev{with a negative scattering length ($a_s<0$)}   is described by the Gross-Pitaevskii equation (GPE) for the macroscopic order parameter $\Psi$:
\begin{align}
\label{GPE}
    i\partial_t \Psi=H\Psi-|\Psi|^2\Psi, \quad H\equiv -\frac{1}{2}\partial_x^2 +V(x,t).
\end{align}
Here $t$ and $x$ are measured respectively in the units $\hbar/2E_r$ and $\Lambda/\pi$, $E_r=\hbar^2\pi^2/(2\Lambda^2 m)$ is the recoil energy, $m$ is the atomic mass, $\Lambda$ is a characteristic length defining the period of the potential $V(x,t)$ whose amplitude is measured in $2E_r$ units, and for a BEC in a cigar-shaped trap with the frequency of the transverse trap $\omega_\bot$ the dimensionless $\Psi$ and dimensional $\Phi$ order parameters are related by  $\Psi=\sqrt{\hbar\omega_\bot|a_s|/E_r}\Phi$    (see e.g.~\cite{Perez}). The dynamical  superlattice is  modeled by 
\begin{align}
\label{pot}
    V(x,t)=-p_1\cos^2(\pi x/d_1)-p_2\cos^2(\pi x/d_2-\nu t)
\end{align}
where $p_{1,2}$ and $d_{1,2}$ are the dimensionless depths \rerev{(in the units of $2E_r$)} and periods of the constitutive lattices. The first stationary lattice can be created by two counter-propagating monochromatic laser beams~\cite{Morsch,BEC1,BEC2}, while the second lattice moving with the dimensionless velocity $v_{\rm L}=\nu d_2/\pi$ is created by two counter-propagating beams with the frequency detuning $\sim \nu$~\cite{shaking,lensing}. We require $v_{\rm L}\ll 1$ to be a small parameter determining adiabatic displacement of the second lattice. \rerev{Considering a $^7$Li BEC ($a_s\approx-1.43\,$nm)
~\cite{Salomon,Hullet}, in a trap with $\omega_\bot=2\pi \times 710$ Hz and $\Lambda\approx 2\,\mu$m,  the period $T=10\pi$ and the distance $L=1$ used in our simulations correspond
to $4.2\,$ms and $0.64\,\mu$m. For a soliton with $10^3$ atoms in such a trap $N=\int_{-\infty}^{\infty} |\Psi|^2dx\approx 8.5$.} The periods $d_1$ and $d_2$ are considered commensurate, i.e., $d_1/d_2=n_1/n_2$ with $n_{1}$ and $n_2$ being coprime integers. At any instant of time the potential remains periodic with the dimensionless period \rev{ $L=n_1d_2=n_2d_1$}.
 Equation (\ref{GPE}) is considered subject to zero boundary conditions at the infinity and the initial conditions ensuring desirable filling of the bands as discussed below. 


\begin{figure*}[htb]
\centering
\includegraphics[width=\textwidth]{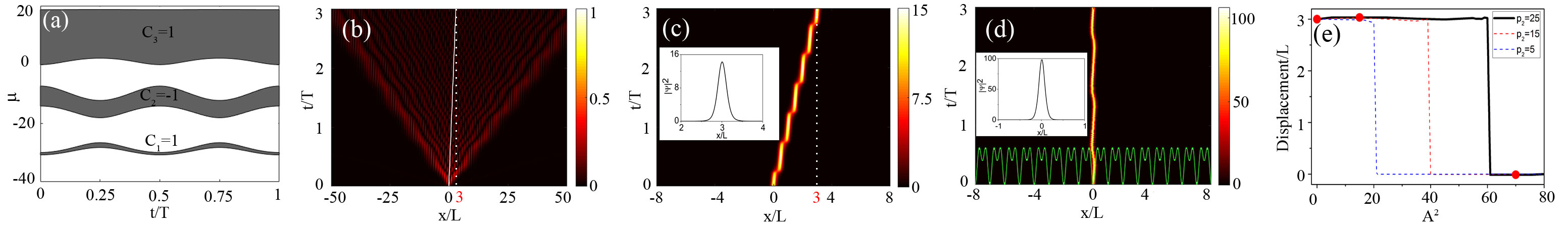}
\caption{ (a) The one-cycle evolution of the first three bands and their Chern numbers. Evolution of the initial state $\Psi_{\rm ini}$ with $\ell=0.4$ 
at $A^2$=0.1 (b), 15 (c) and 70 (d). Solid white curve in (b) visualizes the COM trajectory of the dispersive wavepacket in the quasi-linear regime. Solid green curve in (d) represents the potential profile at $t=0$. Insets in (c) and (d) show the atomic densities at $t=3T$. The vertical dotted lines in (b) and (c) indicate the position of COM at $t=3T$. (e) COM displacement in the dynamical lattices with $p_2=5,\, 15$, and $25$. The red dots on the solid line ($p_2=25$) correspond to dynamics in (b), (c), and (d). In (a-d), $p_1=p_2=25$. In all cases $d_2=2d_1=1$, $L=1$, $\nu=0.1$, and $T=10\pi$.
}
\label{fig:one}
\end{figure*}

Let us consider the coordinate of the COM of the wavepacket defined as $X(t)=N^{-1}\int_{-\infty}^{\infty}x|\Psi|^2dx$.
Adiabatic evolution makes it meaningful to consider instantaneous spectrum of $H$, \rev{i.e., $ H\varphi =\mu  \varphi$ with $\mu$ being a chemical potential} and $t$ considered as a parameter. \rev{ This gives origin to  instantaneous band-gap spectrum $\mu_{\alpha }(k,t)$ [illustrated in Fig.~\ref{fig:one} (a)] with corresponding Bloch functions $\varphi_{\alpha k}(x,t)=e^{ikx} u_{\alpha k}(x,t)$, where $u_{\alpha k}(x,t)=u_{\alpha k}(x+L,t)$, $\alpha=1,2,...$ is the band index, and $k$ is the Bloch momentum.}
 Introducing also the Wannier functions~\cite{Kohn,review_Wannier}, $w_{\alpha n}(x,t)
={\frac{L}{2\pi}}\int_{\text{BZ}}\varphi_{\alpha k}(x,t)e^{-iknL}dk$, we can expand the order parameter $
\Psi
=\sqrt{N}\sum_{\alpha, n} a_{\alpha n} w_{\alpha n}$. The time-dependent coefficients $a_{\alpha n}(t)$ satisfy the normalization condition $\sum_{\alpha,n}|a_{\alpha n}|^2=1$ expressing the fact that norm $N$ is conserved and can be found from the original evolution equation (\ref{GPE})~\cite{supplemental}. This representation allows one to express the shift of the COM in the form 
\begin{align}
    \label{COM-general}
    X(t)=&\rev{\sum_{\alpha=1}^\infty} \rho_\alpha X_\alpha +\rev{L}\rev{\sum_{n=-\infty}^\infty} n\eta_n  
    \nonumber \\
    &+ \rev{\sum_{\alpha,\alpha'=1 \atop \alpha\neq \alpha'}^\infty\sum_{n,n'=-\infty \atop n\neq n'}^\infty} a_{\alpha'n'}^*a_{\alpha n}\int_{-\infty}^{\infty}     w_{\alpha'k'}^*x w_{\alpha k} dx.
\end{align}
Here $\rho_\alpha(t)=\sum_n |a_{\alpha n} |^2$ is the population of the $\alpha$-th band,  
\begin{align}
\label{Xalpha}
 	X_\alpha (t)=\int_{-\infty}^{\infty}w_{\alpha 0}^*(x)x w_{\alpha 0}(x)dx=\frac{L}{2\pi}
 	\int_{\rm BZ} 
 	  \cA_{\alpha} dk 
 \end{align}
describes the transport carried by $\alpha$-th band, $\cA_{\alpha}(t)=\langle u_{\alpha k}(x,t)| i\partial_k   u_{\alpha k}(x,t) \rangle$ is the Berry connection in the $(x,t)$ space, 
$\langle f| g\rangle:=\int_{0}^Lf^*(x)g(x) dx$, $\eta_n(t)=\sum_\alpha |a_{\alpha n} |^2$ is the population of the $n$-th site 
and the last term in Eq.~(\ref{COM-general}) describes contribution of  inter-band transitions.

\paragraph{One-band approximation at weak nonlinearities.} Consider variation of a gapped spectrum over one cycle as illustrated in Fig.~\ref{fig:one}(a) for potential (\ref{pot}). Suppose also that only the lowest band, $\alpha=1$, is populated. Then the COM coordinate is given by a simple expression $X(t)\approx X_1(t)+ \xi_{d}(t)$, where $\xi_d=\rev{L}\sum_{n} n |a_{1n}|^2 $. The pumping term $X_1(t)$ has topological nature. Over one cycle of periodic driving one obtains~\cite{Thouless,ThoulessQH,MiChaNiu2010} $X_1(T)=C_1L$, where
  \begin{align}
  \label{Chern}
      C_\alpha=\frac{i}{2\pi}\int_0^Tdt\int_{BZ}dk\left[\langle \partial_tu_{\alpha k}| 
 \partial_k   u_{\alpha k} \rangle-\langle \partial_k u_{\alpha k}| 
 \partial_t   u_{\alpha k} \rangle
      \right]
  \end{align}
is the Chern number of the band $\alpha$ in $(x,t)~$space~\rev{\cite{supplemental}}. While $X_1(t)$ is determined by the global properties of the lattice, the contribution of the dispersion, $\xi_d(t)$, is determined by the tunneling between the nearest potential minima. In sufficiently deep lattices one  estimates~\rev{\cite{supplemental}}: $|dX_1/dt|\gg |d\xi_d/dt| $. This estimate is valid for both   strong dispersion [Fig.~\ref{fig:one}(b)] and  solitonic [Fig.~\ref{fig:one}(c)] regimes. In any of these limits the \rev{COM coordinate}  being determined by the Chern number of the lowest band is topological.

 Equation~(\ref{GPE}) allows families of soliton solutions bifurcating from the linear Bloch states~\cite{KonSal}. Small-amplitude solitons, $ {A}\to 0$, are characterized by the scaling $N\sim  {A}$.  However, not any soliton input state results in soliton creation. In the absence of a potential, i.e., at $V(x,t)\equiv 0$, Eq.~(\ref{GPE}) is integrable and one can show~\cite{Novikov} that for all $\Psi(x,0)$ satisfying the condition $\int|\Psi(x,0)|dx<U$, where $U=\ln(2+\sqrt{3})$, no solitons are created. At $V(x,t)\neq 0$ a similar estimate is unknown, but one can use the fact that evolution of small-amplitude solitons is well represented by a Bloch wave modulated by slowly varying envelope, ${\mfA}(x,t)$, which is also governed by the NLSE with the effective mass $m_{\rm eff}$ and effective nonlinearity $g_{\rm eff}$ determined by the specific form of the potential~\cite{KonSal}: $ i {\mfA}_t =-(1/2m_{\rm eff}){\mfA}_{xx}-g_{\rm eff}|{\mfA}|^2{\mfA}$. Therefore, in the presence of the lattice the above condition for the absence of solitons (roughly) should include a renormalized constant: $\int|{\mfA}(x,0)|dx\lesssim \sqrt{m_{\rm eff}g_{\rm eff}}U$. Thus, one can distinguish the quasi-linear transport carried by gradually dispersing (non-solitonic) wavepackets [Fig.~\ref{fig:one}(b)] and solitonic transport for which one can observe at least two different regimes, as illustrated in Figs.~\ref{fig:one}(c) and (d).

We numerically solved Eq.~(\ref{GPE}) with the initial excitation condition $\Psi_{\rm ini}(x)=Ae^{-x^2/\ell^2}\varphi_{1,0}(x,0)$, where $\varphi_{1,0}(x,0)$ is the normalized Bloch state of the lowest band at $k=0$, $\ell$ is the width of the envelope, and the amplitude $A$ is used as a control parameter. Increasing $A$ allows us to study the pumping under the transition between the quasi-linear and strongly nonlinear regimes. Quasi-linear transport is shown in Fig.~\ref{fig:one}(b), where one observes that in spite of strong dispersion, the COM of the wavepacket follows the averaged trajectory $X_{\rm av}(t)=Lt/T$,  as predicted above for the lowest band characterized by the first Chern number $C_1=1$. When amplitude of the initial state increases and nonlinearity becomes sufficiently strong \rev{($A>2$ in our simulations)} for compensation of dispersion, a matter-wave soliton forms. Remarkably, its pumping is still topological and is accurately predicted by the formula for $X_1(t)$ [Fig.~\ref{fig:one}(c)]. 


\begin{figure}[htb]
\centering
\includegraphics[width=\columnwidth]{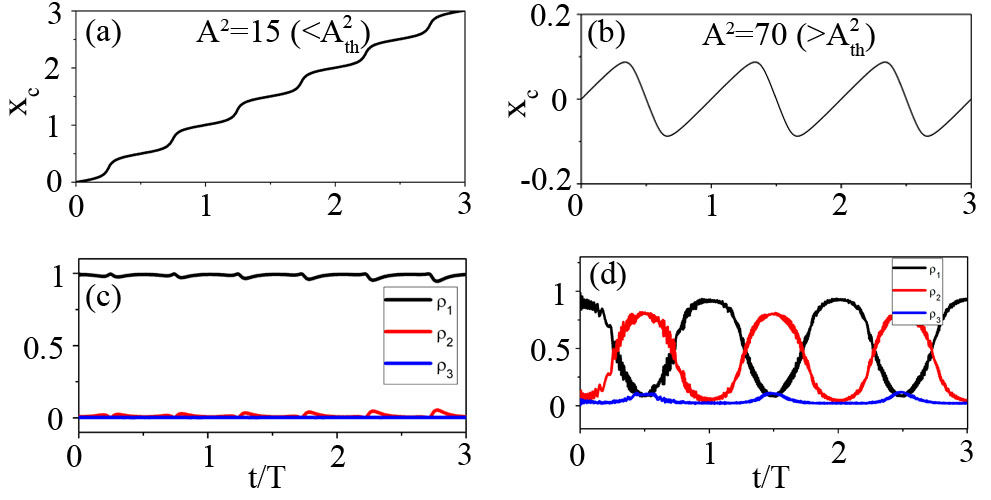}
\caption{(a, b) The displacement of the COM  predicted by the perturbation theory, for the initial amplitude $A$ below (a) and above (b) the threshold amplitude $A_{\rm th}=\sqrt{61}$. (c, d) The evolution of the population of the lowest three bands corresponding to panels (a) and (b). In all cases, $p_1=p_2=25$, $\nu=0.1.$
}
\label{fig:two}
\end{figure}

\paragraph{Solitons and transport breakdown.} Upon further increase of the initial wavepacket amplitude, at some threshold value $A_{\rm th}$ the average forward motion of the soliton becomes inhibited and  small oscillations of a trapped soliton appear [Fig.~\ref{fig:one}(d)].
This phenomenon resembles the dynamical localization in a periodically oscillating potential~\cite{DunKen}, which for solitons  was predicted in~\cite{KCV,BisSal}. There are however two crucial differences indicating on distinct physics of the pumping breakdown. First, the dynamical localization is a linear phenomenon, and the nonlinearity is usually an obstacle for its observation~\cite{BKS}
while trapping shown in Fig.~\ref{fig:one}(d) is a purely nonlinear effect disappearing below threshold amplitude $A_{\rm th}$, that is illustrated in Fig.~\ref{fig:one}(e) for different choices of the lattice parameters. Second, the dynamical localization occurs at a discrete set of frequencies, while the present effect is obtained at all frequencies $\nu$ that guarantee adiabatic variation of the potential. 
To explain breakdown of pumping we recall, that when soliton forms, its expansion via Wannier functions necessarily involves contributions from the upper bands~\cite{KonSal}. Moreover, the larger the nonlinearity the smaller the relative effective depth of the periodic potential, i.e., a larger number of bands are excited. Formation of a soliton from an input pulse occurs over short times, as compared to the long-time adiabatic evolution. Hence, the pulse dynamics can be viewed as a motion of a matter soliton. If the amplitude of the initial pulse is large enough to create a soliton, but not enough to induce the decay of the initial pulse into several solitons~\cite{Satsuma}, the motion of the created soliton can be described using the perturbation theory. 

To this end we consider the renormalized order parameter $\Phi=e^{-i(p_1+p_2)t/2}\Psi/A$, where $A=\sup_x\Psi_{\rm ini}(x) $ is the amplitude of the initial wavepacket (in our simulations of the solitonic regime it is approximately equal to the amplitude $A$ of $\Psi_{\rm ini}$). For sufficiently large $A$, ensuring that $\epsilon=p_1/(2A^2)\ll 1$, the equation for $\Phi$ acquires the form of the perturbed NLSE 
 \begin{equation}
 	\label{NLS-1}
 	i\Phi_\tau+\frac{1}{2} \Phi_{XX}+|\Phi|^2\Phi=-\epsilon \cV(X,\tau)\Phi 
 \end{equation}
where we introduced the scaled variables $\tau=A^2 t $ and $X=A x$, and potential $\cV=\cos(\kappa_1 X)\rev{+}p\cos(\kappa_2X-\omega \tau)$ with $p=p_2/p_1$, $\omega=2\nu/A^2$, and $\kappa_j=2\pi/(A d_j)$ ($j=1,2$).  
Let us choose the initial wavepacket in Eq.~(\ref{NLS-1}) in the form of a soliton of the unperturbed ($\epsilon=0$) NLSE,  
$
    	\Phi_s(X,0)=
    	\lambda e^{i v z/\lambda +i\theta}\mbox{sech} (z), 
$ where $    	
    	 z=\lambda[X-X_c(\tau)]
$,
i.e., we replace $\Psi_{\rm ini}$ by an exact soliton of the same amplitude (even so crude approximation gives accurate prediction of the threshold amplitude). 
The parameters $\lambda$, $v$, and $\theta$ determine the amplitude, velocity, and phase of the soliton, while $X_c(\tau)$ is the COM coordinate (in the unperturbed case  $dX_c/d\tau=v$). 
When $\cV\neq 0$, due to smallness of $\epsilon$, it be considered as perturbation in Eq.~(\ref{NLS-1}), and using the perturbation theory for the NLSE solitons~\cite{perturb}, one obtains $\lambda=$const and
\begin{align}
	\label{eq:xi}
	\frac{d^2X_c}{d\tau^2}=- \epsilon f_1 \sin\left(\kappa_1 X_c\right)-\epsilon p f_2 \sin\left(\kappa_2 X_c-\omega \tau \right) 
\end{align}
where $f_j=\int_{-\infty}^{\infty}\sinh(z)\mbox{sech}^3(z)\sin\left( \kappa_jz/\lambda\right) dz$.   

Figures \ref{fig:two}(a) and (b) show the results of the numerical solution of Eq.~(\ref{eq:xi}) for two different initial amplitudes and $\lambda=1$. In panel (a) we observe the dynamics prescribed by the topological pumping, i.e., by $X_1 (t)$. However, At $A>A_{\rm th}\approx\sqrt{61}$ the wavepacket undergoes periodic oscillations [Fig.~\ref{fig:two}(b)]. This prediction, based on the perturbation theory for solitons, perfectly matches the results of direct numerical simulations of Eq.~(\ref{GPE}) shown by the solid line in Fig.~\ref{fig:one}(e). Predictions of the perturbation theory for threshold amplitude in shallower lattices [see the dashed lines in Fig.~\ref{fig:one}(e)] are also close to results of simulations of Eq.~(\ref{GPE}), but are less accurate because smaller $A_{\rm th}$ corresponds to larger 
$\epsilon$.

 
 \begin{figure}[htb]
\centering
\includegraphics[width=\columnwidth]{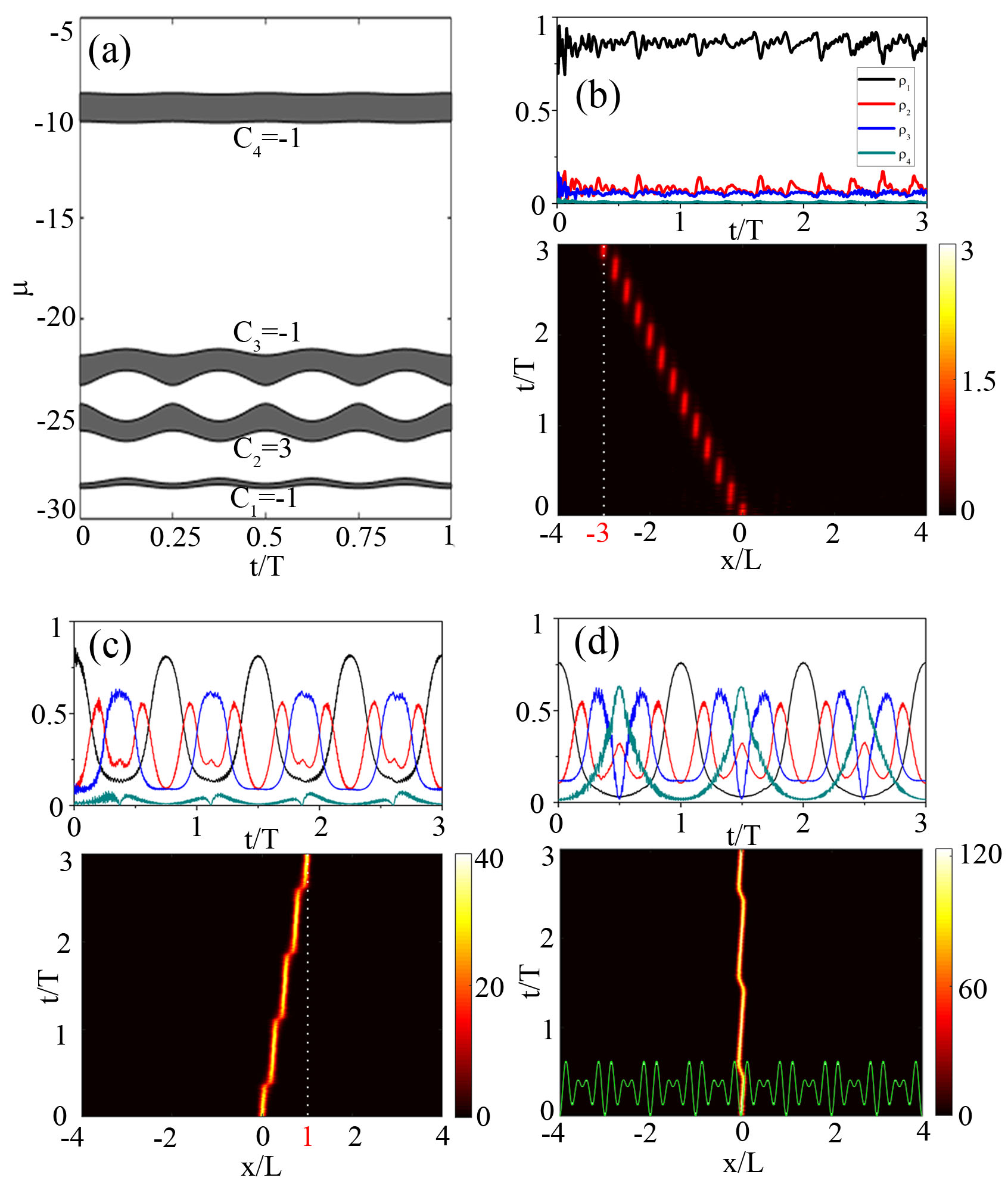}
\caption{(a) The one-cycle evolution of the four lowest bands and their Chern numbers.  (b-d) Evolution of the wavepacket $\Psi_{\rm ini}(x)$ with $
\ell=0.16$ at $A^2=$3 (b), 40 (c) and 100 (d). Panel (c) illustrates fractional pumping. The upper panels in (b)--(d) show atomic populations of the four lowest bands (shown by black, red, blue, and green lines, respectively). The vertical dashed lines in (b,c) indicate the position of the COM at $t=3T$. The solid green curve in the lower panel of (d) represents the potential profile at $t=0$. In all cases, $p_1=p_2=25$, $d_1=1/2$, $d_2=2/3$, and $\nu=0.1$.
}
\label{fig:three}
\end{figure}

\paragraph{Rabi oscillations and topology of the breakdown.}
Remarkably, even the broken pumping can still be interpreted as a topological phenomenon. Indeed, when analysing populations of the lowest bands for unbroken pumping in Fig.~\ref{fig:two}(c), one observes that one-band approximation is perfectly justified: only the first band is populated (black line) and population fluctuates only slightly with time. The situation is dramatically different at $A>A_{\rm th}$: in Fig.~\ref{fig:two}(d) we observe Rabi oscillations of atoms between two lowest bands (black and red lines), while the population of the third band remains, in fact, negligible. Obviously, such situation may be well described with two-band approximation in Eq.~(\ref{COM-general}). Recalling that the Chern number of the second band is $C_2=-1$ [Fig.~\ref{fig:one} (a)], one can interpret spatial oscillations of the soliton in Fig.~\ref{fig:one}(d) and Fig.~\ref{fig:two}(b) as Rabi oscillations between the Bloch bands carrying topological transport in opposite directions. Then forward and backward motion of the soliton corresponds to dominating populations of the first and second bands.
 
To illustrate the interplay of the band topology with inter-band Rabi oscillations, we consider a lattice with Chern numbers of the lowest two bands having different signs and absolute values: $C_1=-1$ and $C_2=3$ [Fig.~\ref{fig:three}(a)]. Focusing on solitonic regime and using the same initial condition as in Fig.~\ref{fig:one}, for relatively small amplitudes in Fig.~\ref{fig:three}(b) we observe Thouless pumping of a soliton predicted by the formula $X(T)=-L$. Since the second and third bands are relatively close, for larger initial amplitudes one excites three lowest bands [Fig.~\ref{fig:three}(c)]. Now the position of the soliton COM is determined by the first and third terms in formula (\ref{COM-general}) describing, respectively, topological pumping and Rabi oscillations among the bands [upper panel of Fig.~\ref{fig:three}(c)]. Now the pumping is fractional and occurs in the opposite (positive) direction of the $x$-axis. The one-cycle-averaged displacement is given by $X(T)=L/3$, and it is relatively well captured by a crude approximation $X_{\rm app}(T)=\sum_{\alpha=1}^3\rho_{\alpha} C_\alpha L$ averaging the contribution of the Rabi-oscillations to zero (for the result shown in Fig.~\ref{fig:three}(c) one has $X_{\rm app}(T)\approx 0.23 L$).
 ~Finally, taking into account that $\sum_{\alpha=1}^4C_\alpha=0$, in order to breakdown the soliton pumping we further increase the amplitude and this indeed leads to the trapping of the soliton shown in Fig.~\ref{fig:three}(d), consistent with the explanation of COM oscillations by Rabi oscillations of the atomic density among four lowest bands.

 
To conclude, we have shown that Thouless pumping preserves its topological nature when considered in soliton bearing nonlinear systems. Recently (after this Letter was submitted) an independent experimental observation of this phenomenon in the array of optical waveguides governed by a discrete nonlinear Schr\"odinger equation was presented~\cite{Rechtsman}. Thus, it is well established that solitons may play a dominant role in transport. Meantime, the present Letter has uncovered physical mechanisms responsible for the quantized transport by solitons in continuous system. We have shown that nonlinear pumping is determined by the well-defined linear Chern numbers of the bands, i.e. the nonlinear evolution is intimately related to the linear topology of the lattice. The role of the nonlinearity in this process is two-fold: it results in the formation of solitons (in the discrete case this is shown in~\cite{Rechtsman}), and it also induces inter-band Rabi oscillations resulting in dynamically varying band populations, which, in turn, determine eventual contributions of higher bands (all accounted by the continuous model) to the soliton displacement. This physical picture allowed us to predict and confirm numerically that by changing only the initial wavepacket, the pump can be reversed with respect to the direction of motion of the dynamical sub-lattice, that it can become fractional or broken (only the latter regime is reported in ~\cite{Rechtsman}). Our results  are applicable for both deep lattices, for which the tight-binding regime considered in~\cite{Rechtsman} is valid, as well as for  relatively shallow lattices requiring fully continuous description. We also have shown that quantized soliton motion in a certain range of amplitudes is well described by the perturbation theory for solitons.

\acknowledgements

V.V.K. acknowledges financial support from the Portuguese Foundation for Science and Technology (FCT) under Contracts PTDC/FIS-OUT/3882/2020 and UIDB/00618/2020. Q.F., P.W. and F.Y. acknowledge support from NSFC (No.91950120,11690033), Natural Science Foundation of Shanghai  (No.19ZR1424400), and Shanghai Outstanding Academic Leaders Plan (No. 20XD1402000).

\clearpage


\begin{thebibliography}{99}


\bibitem{Thouless}   D. J. Thouless,  Quantization of particle transport, {Phys. Rev. B} {\bf 27}, 6083 (1983).

\bibitem{LosZilAlBlo2020} M. Lohse, S. Schweizer, O. Zilberberg, M. Aidelsburger, and I. Bloch,  A Thouless quantum pump with ultracold bosonic atoms in an optical superlattice,  Nat. Phys. {\bf 12}, 350 -- 354 (2016).

\bibitem{LoSHa2018} M. Lohse, C. Schweizer, H. M. Price, O. Zilberberg, and I. Bloch, Exploring 4D quantum Hall physics with a 2D topological charge pump, Nature {\bf 553}, 55
(2018). 

\bibitem{NaTaKe2021} S. Nakajima, N. Takei, K. Sakuma, Y. Kuno, P. Marra, and  Y. Takahashi,
Competition and interplay between topology and quasi-periodic disorder in Thouless pumping of ultracold atoms. Nat. Phys.  {\bf 17}, 844 (2021).

\bibitem{NaToTa2016} S. Nakajima, T. Tomita, S. Taie, T. Ichinose, H.  Ozawa, L. Wang, M. Troyer, and Y. Takahashi, Topological Thouless pumping of ultracold fermions.  Nat. Phys. {\bf 12}, 296 
(2016).

\bibitem{spin} 
W. Ma, L. Zhou, Q. Zhang, M. Li, C. Cheng, J. Geng, X. Rong, F. Shi, J. Gong, and J. Du, Experimental Observation of a Generalized Thouless Pump with a Single Spin, Phys. Rev. Lett. {\bf 120}, 120501 (2018).

\bibitem{KraLaRin2012} Y. E. Kraus,  Y. Lahini, Z. Ringe, M. Verbin, and  O. Zilberberg, 
Topological States and Adiabatic Pumping in Quasicrystals,  { Phys Rev. Lett.} {\bf 109}, 106402 (2012).

\bibitem{ZilHuaJo2018} O. Zilberberg, S. Huang, J. Guglielmon, M. Wang, K. P. Chen, Y. E. Kraus, and M. C. Rechtsman, Photonic topological boundary pumping as a probe of 4D quantum Hall physics,  Nature, {\bf 553}, 59 (2018).

\bibitem{CeWaShe2020} A. Cerjan, M. Wang,  S. Huang, K. P. Chen, and M. Rechtsman, Thouless pumping in disordered photonic systems, Light: Science \& Applications {\bf 9}, 178 (2020).


\bibitem{CheProPro2020} 
W. Cheng, E. Prodan, and C. Prodan, 
Experimental Demonstration of Dynamic Topological Pumping across Incommensurate Bilayered Acoustic Metamaterials, Phys. Rev. Lett. {\bf 125}, 224301 (2020).
 

%


\bibitem{FedQiLIK2020} Z. Fedorova, H. Qiu, S. Linden, and  J. Kroha,  
Observation of topological transport quantization by dissipation in fast Thouless pumps,
{Nat. Comm.} {\bf 11}, 3758 (2020).

\bibitem{MiChaNiu2010} D. Xiao, M.-C. Chang, and Q.  Niu,  Berry phase effects on electronic properties, Rev. Mod. Phys. {\bf 82}, 1959 (2010).

\bibitem{LuSAG2016}  H.-I. Lu, M. Schemmer, L. M.  Aycock, D. Genkina, S. Sugawa,  and I. B. Spielman, Geometrical Pumping with a Bose-Einstein Condensate, Phys. Rev. Lett. {\bf 116}, 200402 (2016).

\bibitem{TaCoFa2017}
L. Taddia, E. Cornfeld, D. Rossini, L. Mazza, E. Sela, and R. Fazio, Topological Fractional Pumping with Alkaline-Earth-Like Atoms in Synthetic Lattices, Phys. Rev. Lett. {\bf 118}, 230402 (2017).


\bibitem{HaDuKAm2019} T Haug, R. Dumke, L.-C. Kwek, and L. Amico, Topological pumping in Aharonov -- Bohm rings, Comm. Phys. {\bf 2}, 127 (2019).

\bibitem{TanDasAn2016} J. Tangpanitanon, V. M. Bastidas, S. Al-Assam, P. Roushan, D. Jaksch, and D. G. Angelakis, Topological Pumping of Photons in Nonlinear Resonator Arrays, Phys. Rev. Lett. {\bf 117}, 213603 (2016).

\bibitem{NaYoKa2018} M. Nakagawa, T. Yoshida,  R. Peters,  and N. Kawakami, Breakdown of topological Thouless pumping in the strongly interacting regime,
Phys. Rev. B {\bf 98}, 115147 (2018).

\bibitem{SteHaHe2019} L. Stenze, A. L. C. Hayward, C. Hubig, U. Schollw\"ock, and F. Heidrich-Meisner, Quantum phases and topological properties of interacting fermions in one-dimensional superlattices, Phys. Rev. A {\bf 99}, 053614 (2019).

\bibitem{AKKS} G. L. Alfimov, P. G. Kevrekidis, V. V. Konotop, and M. Salerno, Wannier functions analysis of the nonlinear Schr\"odinger equation with a periodic potential, Phys. Rev. E {\bf 66}, 046608 (2002). 

\bibitem{BraKon2004} V. A. Brazhnyi and V. V. Konotop, Theory of nonlinear matter waves in optical lattices", Mod. Phys. Lett. {\bf 18}, 627 (2004).

\bibitem{perturb} V. I. Karpman and E. M. Maslov, Perturbation theory for solitons, Sov. Phys. JETP {\bf 48}, 252 (1978).

\bibitem{KCV} V. V. Konotop, O. A. Chubykalo, and L. V\'azquez, Dynamics and interaction of solitons on an integrable inhomogeneous lattice, Phys. Rev. E {\bf 48}, 563 (1993).

\bibitem{BisSal} D. Cai, A. R. Bishop, N Gr\o nbech-Jensen, and M. Salerno,
Electric-Field-Induced Nonlinear Bloch Oscillations and Dynamical Localization, 
Phys. Rev. Lett. {\bf 74}, 1186 (1995).  

\bibitem{optlatt01} Y. V. Kartashov, L. Torner, and D. N. Christodoulides, Soliton dragging by dynamical optical lattices, Opt. Lett. \textbf{30}, 1378 (2005).

\bibitem{optlatt02} I. L. Garanovich, A. A. Sukhorukov, and Y. S. Kivshar, Soliton control in modulated optically-induced photonic lattices, Opt. Express \textbf{13}, 5704 (2005).

\bibitem{optlatt03} C. R. Rosberg, I. L. Garanovich, A. A. Sukhorukov, D. N. Neshev, W. Krolikowski, and Y. S. Kivshar, Demonstration of all-optical beam steering in modulated photonic lattices, Opt. Lett. \textbf{31}, 1498 (2006).

\bibitem{zhangoe2015} X. Zhang, F. Ye, Y. Kartashov, and X. Chen, Rabi oscillations and simulated mode conversions on the subwavelength scale, Opt. Express \textbf{23}, 6731 (2015).

\bibitem{BEC1} H. Lignier, C. Sias, D. Ciampini, Y. Singh, A. Zenesini, O. Morsch, and E. Arimondo, Dynamical Control of Matter-Wave Tunneling in Periodic Potentials, Phys. Rev. Lett.  {\bf 99}, 220403 (2007).

\bibitem{BEC2} A. Zenesini, H. Lignier, C. Sias, O. Morsch, D. Ciampinia, and E. Arimondo, 
Tunneling Control and Localization for Bose--Einstein Condensates 
in a Frequency Modulated Optical Lattice, Las. Phys, {\bf 20}, 1182
(2010).

\bibitem{StaLon} K. Staliunas and S. Longhi, Subdiffractive solitons of Bose-Einstein condensates in time-dependent optical lattices, Phys. Rev. A {\bf 78}, 033606 (2008).

\bibitem{Perez} V. M. P\'erez-Garc\'ia, H. Michinel, and H. Herrero, Bose-Einstein solitons in highly asymmetric traps. Phys. Rev. A {\bf 57}, 3837 (1998)


\bibitem{Morsch} O. Morsch and M. Oberthaler, 
Dynamics of Bose-Einstein condensates in optical lattices.
Rev. Mod. Phys. {\bf 78}, 179 (2006).

\bibitem{shaking}
H. Lignier, C. Sias, D. Ciampini, Y. Singh, A. Zenesini, O. Morsch, and E. Arimondo, Dynamical Control of Matter-Wave Tunneling in Periodic Potentials. Phys. Rev. Lett. {\bf 99}, 220403 (2007).

\bibitem{lensing}
L. Fallani, F. S. Cataliotti, J. Catani, C. Fort, M. Modugno, M. Zawada, and M. Inguscio, Optically Induced Lensing Effect on a Bose-Einstein Condensate Expanding
in a Moving Lattice. Phys. Rev. Lett. {\bf 91}, 240405 (2003).

\bibitem{Salomon} \rerev{L. Khaykovich, F. Schreck, G. Ferrari, T. Bourdel, J. Cubizolles, L. D. Carr, Y. Castin, and C. Salomon,	Formation of a Matter-Wave Bright Soliton Science, {\bf 296}, 1290 (2002).}

\bibitem{Hullet} \rerev{K. E. Strecker, G. B. Partridge, A. G Truscott, and R. G. Hulet, Formation and propagation of matter-wave soliton trains, Nature {\bf 417}, 150 (2002).}

\bibitem{Kohn} W. Kohn, Analytic Properties of Bloch Waves and Wannier Functions, Phys. Rev. {\bf 115}, 809 (1959).

\bibitem{review_Wannier} Marzari, N.,   Mostof, A. A.,   Yates,  J. R.,  Souza, I. \& Vanderbilt, D., Maximally localized Wannier functions: Theory and applications,  {Rev. Mod. Phys.} {\bf 84}, 1419 (2012).


\bibitem{supplemental} See the details in the Supplemental Materials, which includes Refs. [19,37,40,47].
\bibitem{fft} M. D. Feit, and J. A. Fleck, Calculation of dispersion in graded-index multimode fibers by propagting beam method, Appl. Opt. {\bf 18},2843 (1979).

\bibitem{ThoulessQH} D. J. Thouless, M. Kohmoto,  M. P. Nightingale,   and M. Nijs,  Quantized hall conductance in a two-dimensional periodic potential. Phys. Rev. Lett. {\bf 49}, 405
(1982).

 

\bibitem{KonSal} V. V. Konotop and M. Salerno, Modulational instability in Bose-Einstein condensates in optical lattices. Phys. Rev. A {\bf 65}, 021602(R) (2002). 

\bibitem{Novikov}  S. Novikov, S. V.  Manakov, L. P.  Pitaevskii, and V. E. Zakharov, Theory of Solitons. The Inverse Scattering Method (Consultants Bureau, New York, 1984).

\bibitem{DunKen} D. H. Dunlap and V. M. Kenkre, Dynamic localization of a charged particle moving under the influence of an electric field, Phys. Rev. B {\bf 34}, 3625 (1986).

\bibitem{BKS} Yu. V. Bludov, V. V. Konotop, and M. Salerno, Dynamical localization of gap-solitons by time periodic forces, EPL {\bf 87}, 20004 (2009).

\bibitem{Satsuma} J. Satsuma and N. Yajima, Initial Value Problems of One-Dimensional Self-Modulation of Nonlinear Waves in Dispersive Media, Suppl. Prig. Theor. Phys.  No. 55, 284 (1974).

\bibitem{Rechtsman} \rerev{M. J\"urgensen, S. Mukherjee, and M. C. Rechtsman, Quantized nonlinear Thouless Pumping, Nature {\bf 596}, 63
(2021).}






%
%


 



 

 



 
\end{thebibliography}
\end{document}